\documentclass[10pt]{article}
\usepackage{graphicx}
\usepackage{amsmath}
\usepackage{amssymb}
\usepackage{color}
\usepackage{caption2}
\begin{document}
\bibliographystyle{prsty}
\begin{center}
{\large {\bf \sc{  Comment on "OPE and quark-hadron duality for two-point functions of tetraquark currents in $1/N_c$
 expansion"  }}} \\[2mm]
Zhi-Gang Wang \footnote{E-mail: zgwang@aliyun.com.  }     \\
 Department of Physics, North China Electric Power University, Baoding 071003, P. R. China
\end{center}

\begin{abstract}
Without excluding the contributions of factorizable Feynman diagrams in the color space to the QCD sum rules by hand, we cannot obtain the conclusion that the factorizable parts of  the operator product expansion series cannot have any relationship to the possible tetraquark bound states. The tetraquark couplings  $f_T$ are  of the order $\mathcal{O}(N_c)$ rather than of the order $\mathcal{O}(N^0_c)$ in the large $N_c$ limit, the  conclusion "a possible exotic tetraquark state may
appear only in $N_c$-subleading contributions to the QCD Green functions" is a paradox.
\end{abstract}

A hadron has many Fock states, a tetraquark state, which has four valence quarks, maybe have color-singlet-color-singlet ($11$) type,   color-antitriplet-color-triplet ($\bar{3}3$) type or
color-sextet-color-antisextet ($6\bar{6}$) type Fock states. If a hidden-charm tetraquark state has the $11$-type Fock states, we can construct the $11$-type four-quark currents, which should have the same  quantum numbers, to interpolate  this  hidden-charm tetraquark state, because the quantum field theory does not forbid such  current-tetraquark couplings.  The argument  applies  to other Fock states.

Now let us construct the four-quark currents to interpolate the hidden-charm tetraquark states,
\begin{eqnarray}
J_{11}(x,\epsilon)&=&\bar{c}(x+\epsilon)\Gamma q(x+\epsilon)\,\bar{q}^\prime(x) \Gamma^\prime c(x)\, ,\nonumber\\
J_{\bar{3}3}(x,\epsilon)&=& \varepsilon^{ijk}\varepsilon^{imn} q^{T}_j(x)C\Gamma c_k(x)  \bar{q}^{\prime }_m(x+\varepsilon)\Gamma^\prime C \bar{c}^{T}_n(x+\varepsilon)\, ,\nonumber\\
J_{6\bar{6}}(x,\epsilon)&=& q^{T}_j(x)C\Gamma c_k(x)  \bar{q}^{\prime }_j(x+\varepsilon)\Gamma^\prime C \bar{c}^{T}_k(x+\varepsilon)+q^{T}_j(x)C\Gamma c_k(x)  \bar{q}^{\prime }_k(x+\varepsilon)\Gamma^\prime C \bar{c}^{T}_j(x+\varepsilon)\, ,
\end{eqnarray}
where the $i$, $j$, $k$, $m$, $n$ are color indexes, the $\Gamma$ and $\Gamma^\prime$ are Dirac $\gamma$-matrixes, the $\epsilon$ is the spatial separation between the two clusters in the color space.

The tetraquark states are spatial extended objects,  if the mean spatial sizes $\langle r\rangle \geq \epsilon$, we can choose the currents $J_{11}(x,\epsilon)$, $J_{\bar{3}3}(x,\epsilon)$ and $J_{6\bar{6}}(x,\epsilon)$ to interpolate the $11$-type,  $\bar{3}3$-type and $6\bar{6}$-type tetraquark states, respectively, although the diquark states in color-sextet is not favored by repulsive interaction originates from the one-gluon exchange.

In fact, it is difficult to take into account the nonlocal effects in the QCD sum rules due to the appearance of the $\epsilon$,  as we have to deal with a bound state problem,   so we usually take the local limit $\epsilon \to 0$. { \textcolor{magenta}{In the local limit $\epsilon \to 0$, the currents $J_{11}(x,0)$ couple potentially to the $11$-type tetraquark states rather than two-meson pairs, because  in such a small spatial separation, the $\bar{c}q$ and $\bar{q}^\prime c$ mesons lose themselves and merge into tetraquark states}}. Direct calculations based on the QCD sum rules support such arguments \cite{WangZG-nonlocal,WangZG-Landau}.

If the mean spatial sizes $\langle r\rangle < \epsilon$, the currents $J_{11}(x,\epsilon)$ couple potentially to the two-meson pairs, because in such large spatial separations, the $\bar{c}q$ and $\bar{q}^\prime c$ mesons retain themselves.

Generally speaking, the  $\bar{3}3$-type and $6\bar{6}$-type tetraquark states can have larger spatial extensions, as the confinement forbids the appearance of the free diquark states \cite{WangZG-nonlocal}.

In the local limit, the currents $J_{\bar{3}3}(x,0)$ and $J_{6\bar{6}}(x,0)$  can be transformed  into the currents   $J_{11}(x,0)$ freely through Fierz rearrangements  in the color and Dirac-spinor  spaces  \cite{WangZG-nonlocal}, we can study the current $J_{11}(x,0)$ as an example.  The $J_{11}(0,0)$, $J_{\bar{3}3}(0,0)$ and $J_{6\bar{6}}(0,0)$ couple potentially to the tetraquark states ($T$),
\begin{eqnarray}
\langle 0|J_{11/\bar{3}3/6\bar{6}}(0,0)|T(p)\rangle &=&f_T\, ,
\end{eqnarray}
where the $f_T$ are the pole residues or decay constants or tetraquark couplings.

\begin{figure}
 \centering
  \includegraphics[totalheight=6cm,width=8cm]{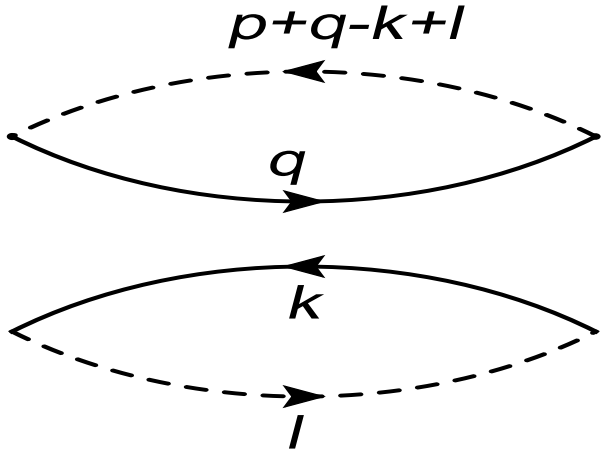}
 \caption{ The  Feynman diagrams  for the lowest order  contributions, where the solid lines and dashed lines represent  the light quarks and heavy quarks, respectively. }\label{Lowest-diagram}
\end{figure}

Now we write down  the two-point correlation functions $\Pi(p)$  in the QCD sum rules,
\begin{eqnarray}
\Pi(p)&=&i\int d^4x e^{ip \cdot x} \langle0|T\Big\{J_{11}(x,0)J_{11}^{\dagger}(0,0)\Big\}|0\rangle \, ,
\end{eqnarray}
The lowest order  Feynman diagrams shown in Fig.\ref{Lowest-diagram} can be divided into two color-neutral clusters, each cluster corresponds to a trace both in the color space and in the Dirac spinor space, the lowest order  Feynman diagrams are proportional to $N_c^2$ in the large $N_c$ limit. In the momentum space, they are nonfactorizable diagrams, the basic integrals are of the form,
\begin{eqnarray}\label{Basic-Integral}
\int d^4q d^4k d^4l \frac{1}{\left(p+q-k+l\right)^2-m_c^2+i\epsilon}\frac{1}{q^2-m_q^{\prime2}+i\epsilon}\frac{1}{k^2-m_q^2+i\epsilon}\frac{1}{l^2-m_c^2+i\epsilon}\, .
\end{eqnarray}
Such integrals certainly have an imaginary part, and we can obtain imaginary parts  through dispersion relation after carrying out the integrals over $q$, $k$, $l$, and they make contributions  to the QCD sum rules. From the Fig.\ref{Lowest-diagram}, we can obtain the relation,
 \begin{eqnarray}
 f_{T}&\propto& N_c\, ,
 \end{eqnarray}
 rather than the relation $f_{T}\propto N_c^0$ \cite{Chush-2021} in the large $N_c$ limit. In Ref.\cite{Chush-2019},  Lucha,  Melikhov and  Sazdjian discard
  the factorizable Feynman diagrams in the color space via putting the confined quark and gluons on the mass-shell and applying Landau equation to study them by hand.
 In Ref.\cite{WangZG-Landau}, we present detailed discussions to show that the Landau equation is of no use  to study  the Feynman diagrams in the QCD sum rules for the tetraquark  states,  the   tetraquark  states begin to receive contributions at the order $\mathcal{O}(\alpha_s^0/\alpha_s^1)$ rather than at the order $\mathcal{O}(\alpha_s^2)$ claimed in Ref.\cite{Chush-2019}. In Ref.\cite{Chush-2021}, Lucha,  Melikhov and  Sazdjian re-express their viewpoint as   the   tetraquark  states begin to receive contributions at the order $\mathcal{O}(\alpha_s^2 N_c^2)$ in the large $N_c$ limit and obtain the  conclusion "a possible exotic tetraquark state may
appear only in $N_c$-subleading contributions to the QCD Green functions", which  is a paradox.

\section*{Acknowledgements}
This  work is supported by National Natural Science Foundation, Grant Number  11775079.

\end{document}